\documentclass[aps,twocolumn,nofootinbib,superscriptaddress,preprintnumbers,floatfix]{revtex4}

\usepackage{graphicx}

\usepackage[hypertex]{hyperref}

\newcommand{\be}{\begin{equation}}
\newcommand{\ee}{\end{equation}}

\newcommand{\ov}{\overline}
\def\d{{\rm d}}
\def\OMIT#1{{}}

\newcommand{\Tr}{\mathrm{Tr}}

\newcommand{\GeV}{~\mathrm{GeV}}
\newcommand{\TeV}{~\mathrm{TeV}}

\newcommand{\pb}{\mathrm{pb}}
\newcommand{\fb}{\mathrm{fb}}

\newcommand{\Br}{\mathcal{B}}

\newcommand{\Eq}[1]{(\ref{#1})}
\newcommand{\Eqs}[2]{Eqs.~(\ref{#1}) and (\ref{#2})}

\def\bentarrow{\raisebox{5pt}{\rlap{$\vert$}}\hspace*{-2.17pt}\to}

\begin{document}

\title{Supermodels for early LHC}

\author{Christian W.\ Bauer}
\affiliation{Theoretical Physics Group, Lawrence Berkeley
  National Laboratory, Berkeley, CA 94720}
\affiliation{Berkeley Center for Theoretical Physics,
  University of California, Berkeley, CA 94720}

\author{Zoltan Ligeti}
\affiliation{Theoretical Physics Group, Lawrence Berkeley
  National Laboratory, Berkeley, CA 94720}
\affiliation{Berkeley Center for Theoretical Physics,
  University of California, Berkeley, CA 94720}

\author{Martin Schmaltz}
\affiliation{Theoretical Physics Group, Lawrence Berkeley
  National Laboratory, Berkeley, CA 94720}
\affiliation{Berkeley Center for Theoretical Physics,
  University of California, Berkeley, CA 94720}
\affiliation{Physics Department, Boston University, Boston, MA 02215}
  
\author{Jesse Thaler}
\affiliation{Theoretical Physics Group, Lawrence Berkeley
  National Laboratory, Berkeley, CA 94720}
\affiliation{Berkeley Center for Theoretical Physics,
  University of California, Berkeley, CA 94720}

\author{Devin G.\ E.\ Walker}
\affiliation{Theoretical Physics Group, Lawrence Berkeley
  National Laboratory, Berkeley, CA 94720}
\affiliation{Berkeley Center for Theoretical Physics,
  University of California, Berkeley, CA 94720}
  \affiliation{Center for the Fundamental Laws of Nature, Jefferson Physical Laboratory,
Harvard University, Cambridge, MA 02138}

\begin{abstract}

We investigate what new physics signatures the LHC can discover in the
2009--2010 run, beyond the expected sensitivity of the Tevatron data by 2010. 
We construct ``supermodels", for which the LHC sensitivity even with only
10~pb$^{-1}$ is greater than that of the Tevatron with 10~fb$^{-1}$.  The
simplest supermodels involve $s$-channel resonances in the quark-antiquark and
especially in the quark-quark channels.  We concentrate on easily visible final
states with small standard model backgrounds, and find that there are simple
searches, besides those for $Z'$ states, which could discover new physics in
early LHC data.  Many of these are well-suited to test searches for ``more
conventional" models, often discussed for multi-fb$^{-1}$~data sets.

\end{abstract}

\maketitle

\section{Introduction}

In this paper, we explore the new physics discovery potential of the first LHC
run, expected to start later this year.  The latest official schedule calls for
7 TeV collisions starting in late 2009 with a ramp-up towards 10 TeV sometime
during the run, which will last until late 2010~\cite{LHCrun}.  However, there
is still some uncertainty in the ultimate center of mass energy, and the useful
luminosity for physics analyses may be significantly less than the
200--300~pb$^{-1}$ delivered luminosity, which is projected for this
run.  We therefore find it interesting to study the sensitivity of the first run
as a function of LHC energy and luminosity.

In particular, it is often stated that a first LHC run with order 10 pb$^{-1}$ 
of good data to be analyzed by \mbox{ATLAS} and CMS would essentially be an
``engineering run" with only the capability to ``rediscover'' the standard
model~\cite{chamonix,koeneke}.  One expects that order 100~pb$^{-1}$ of data will 
be necessary for the LHC to have sensitivity to plausible new physics scenarios.  
Here we take a fresh look at the new physics capabilities of a 10~pb$^{-1}$
low-luminosity data set, and allow ourselves to contemplate new physics which is
not motivated by model building goals such as unification, weak scale dark
matter, or solving the hierarchy problem.

We find that, setting such model building prejudices aside, there is a set
of interesting new physics scenarios that could give rise to a clean, observable
signal in early LHC data, while not being detected with 10~fb$^{-1}$ of Tevatron
data (roughly the projected integrated luminosity at the end of 2010).  These models
are consistent with previous experiments such as LEP II, precision electroweak
constraints, and flavor physics.  Moreover, these scenarios have similar
signatures to ``well-motivated'' new physics models that require higher
luminosity for discovery.  

To set the stage, recall that the production cross sections for new hypothetical
particles can be quite large.  For example, QCD pair production of 500 GeV
colored particles have cross sections in the pb range, such that tens of such
particles could be produced in early LHC.  Of course, in order for the new
particles to be observable, they must have sufficiently large branching
fractions to final states with distinctive signatures and controllable standard
model backgrounds.  Also, the new particles should not be ruled out by current
or future Tevatron searches, implying that the cross section times integrated
luminosity at the LHC should be larger than the corresponding quantity at the
Tevatron.  

Thus, the four criteria for a new physics scenario to be discovered in
early LHC with low luminosity are:
\begin{enumerate}
\item Large enough LHC cross section to produce at least 10 signal
events\footnote{While fewer events might be sufficient for discovery, we shall demand 10 events
to allow for $\mathcal{O}(1)$ uncertainties in our analysis.} with 10~pb$^{-1}$ of
data;
\item Small enough Tevatron cross section to evade the projected 2010 Tevatron
sensitivity with $10\ \fb^{-1}$;
\item Large enough branching fraction to an ``easy'' final state with
essentially no backgrounds;
\item Consistency with other existing bounds.
\end{enumerate}
We call a new physics scenario satisfying these conditions a
\emph{supermodel\/}.

The classic example that comes to mind as a candidate supermodel is a TeV-scale
$Z'$ boson~\cite{Zprime_pdg}. Assuming the $Z'$ mass exceeds the Tevatron reach, but is light
enough and has large enough couplings so that it can be produced copiously at
the LHC, then it can be discovered through its decay to electron and muon
pairs.  Such leptonic finals states are ``easy" to reconstruct with a peak in
the invariant mass distribution, which reduces the already low standard model
backgrounds.

As we will see, however, a typical leptonically decaying $Z'$ is actually not a
supermodel.  First, since the $Z'$ is produced via the quark-antiquark initial
state, the Tevatron is quite competitive with the LHC.  Second, the leptonic
branching fraction is severely bounded by LEP~II data, which restricts the
couplings of the $Z'$ to leptons.  It is therefore nontrivial to find
supermodels that are as discoverable as a standard $Z'$ but consistent with
known bounds on new physics.

The remainder of the paper is organized as follows.  In
Sec.~\ref{sec:production}, we identify new particle production channels with
sufficiently large LHC cross sections and for which the LHC has an advantage
over the Tevatron.  Assuming perturbative couplings, we find that $s$-channel
production of quark-quark ($qq$) or quark-antiquark ($q\bar{q}$) resonances are the best starting points for
early LHC supermodels.  In Sec.~\ref{sec:decay}, we construct
explicit models where these  resonances can decay to interesting and easily
reconstructable final states.  While a standard $Z'$ does not work, generalized
$Z'$ scenarios can be supermodels, as are scenarios involving diquarks.  We
conclude in Sec.~\ref{sec:conclude}.

\section{Production Modes}
\label{sec:production}

\begin{figure}[bt]
\includegraphics[width=\columnwidth]{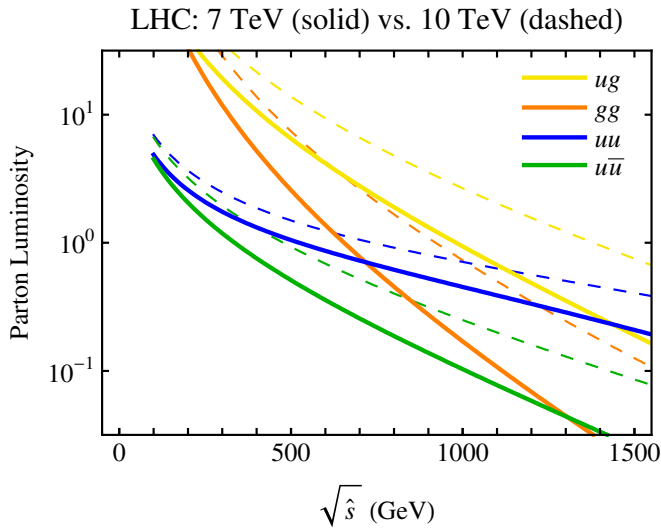}
\caption{LHC parton luminosities as defined in Eq.~\Eq{eq:partonluminosity}, as
functions of the partonic invariant mass.  The solid (dashed) curves are for the
7 TeV (10 TeV) LHC.  The up quark has been chosen as a representative quark, and
each curve includes the contribution from the $CP$ conjugate initial partons.}
\label{fig:rawlhcluminosity}
\end{figure}

In this section, we discuss which production modes have the potential to be
supermodels, deferring detailed model building to Sec.~\ref{sec:decay}. 
Since the expected integrated luminosity at the Tevatron ($\sim
10~\mathrm{fb}^{-1}$) is orders of magnitude larger than our
$10~\mathrm{pb}^{-1}$ benchmark luminosity for early LHC analysis, and since $p
\bar{p}$ parton luminosities are not so different from $p p$ parton
luminosities, one must consider sufficiently heavy new particles to evade the
Tevatron reach. We will find that the most promising perturbative scenarios
accessible with 10~pb$^{-1}$ of LHC data are $qq$ and $q\bar q$ resonances.

\begin{figure}[bt]
\includegraphics[width=\columnwidth]{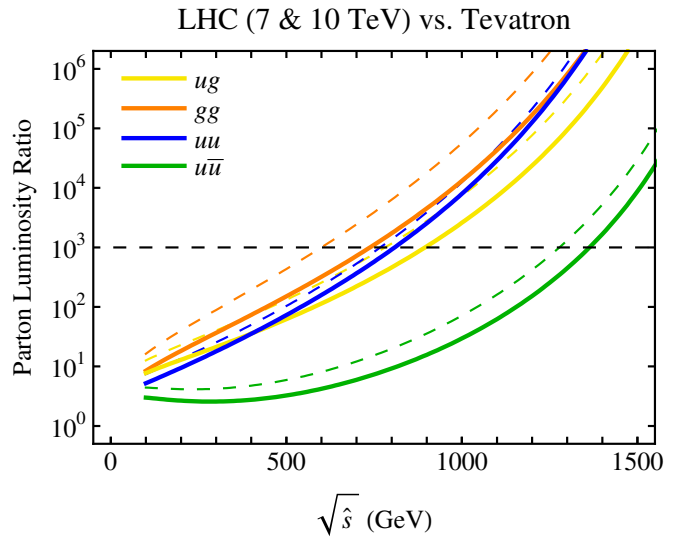}
\caption{Ratios of the parton luminosities for 7 TeV (solid) and 10 TeV (dashed)
LHC compared to the 1.96 TeV Tevatron, as functions of the partonic invariant mass. 
When this ratio is above the $10^3$ horizontal dashed line, the LHC with
10~pb$^{-1}$ will have greater sensitivity than the Tevatron with 10~fb$^{-1}$.}
\label{fig:lumiratios}
\end{figure}

To begin, we plot in
Fig.~\ref{fig:rawlhcluminosity} the LHC parton luminosities, defined as
\be
\label{eq:partonluminosity}
{\cal F}_{ij}(\hat s, s) =
  \int_{\hat s/s}^1\! \d x_i\, \frac{\hat s}{x_i s}\, 
  f_i(x_i)\, f_j[\hat s/(x_i s)]\,,
\ee 
and in Fig.~\ref{fig:lumiratios} the ratios of each parton luminosity at the 
LHC and the Tevatron.  In Eq.~(\ref{eq:partonluminosity}), $\sqrt{s}$ is the
center of mass energy of the collider, $\sqrt{\hat s}$ is the invariant mass of
the two interacting partons, and $f_i(x_i)$ are the parton distribution
functions evaluated at a momentum fraction $x_i$ and scale $\sqrt{\hat s}$.  We
use the CTEQ-5L parton distribution functions~\cite{Lai:1999wy}.  (For similar plots
using CTEQ-6L1 \cite{Pumplin:2002vw}, see Ref.~\cite{Quigg:2009gg}.)

It is often stated that the LHC is essentially a gluon collider, so one might
think that processes initiated by gluons would be the best starting points for
constructing supermodels. However, Fig.~\ref{fig:rawlhcluminosity} shows that the
$gg$ parton luminosity only dominates for small invariant mass, where the
initial LHC data set cannot compete with the Tevatron.  As seen in
Fig.~\ref{fig:lumiratios}, only at large invariant masses do the LHC parton
luminosities become sufficiently enhanced compared to the Tevatron. (The
enhancement of the $q\bar q$ channel is the smallest, so it is harder for the
LHC to compete in cases where the initial $q\bar q$ state contributes.)  To build
supermodels, we must explore the possible LHC cross sections in the region
with large enough enhancements compared to the Tevatron. We will emphasize this
point in the next subsection by showing why QCD pair production is not a
supermodel, and then go on to consider supermodels constructed from $s$-channel
resonances.

\subsection{QCD pair production?}
\label{subsec:nopair}

\begin{figure}[bt]
\includegraphics[width=0.45\textwidth]{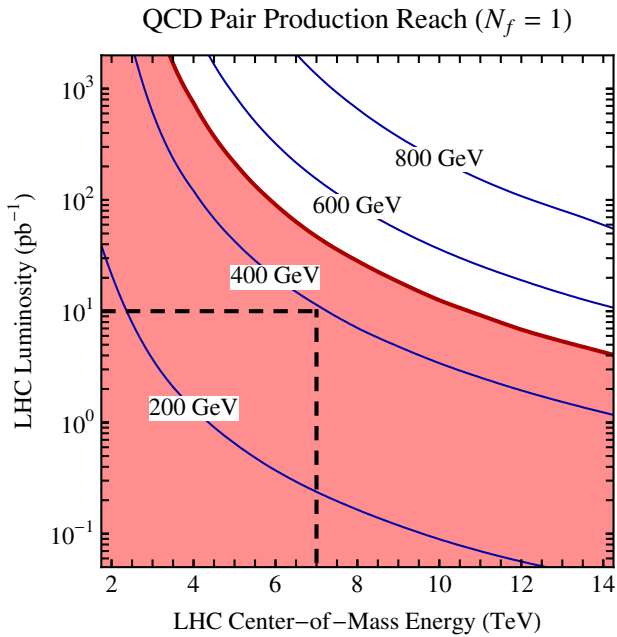}
\caption{LHC reach for pair production of a single flavor of heavy quark as
a function of energy and luminosity.  Each contour corresponds to the production
of 10 events at the LHC for the indicated quark mass.  The red region
corresponds to quark masses which the Tevatron would be able to rule out
with 10~fb$^{-1}$, because it would produce 10 or more events.
The intersection of the straight dashed lines touches the contour corresponding 
to the maximum quark mass ($\sim 400$ GeV) probed by the 7~TeV LHC with 10~pb$^{-1}$ of data.  One sees
that the early LHC is generically not sensitive to QCD pair production of quarks
with masses beyond the Tevatron reach.}
\label{QCDpairs}
\end{figure}

A simple process initiated by gluons is QCD pair production of new colored
particles.  For not too heavy states, it can have a cross section above a pb,
yielding ${\cal O}(10)$ events with 10~pb$^{-1}$ of LHC data. However, it is
easy to show that such processes are generically not supermodels.  For
concreteness, we study the production of a color-triplet quark $Q$.  We assume
that it always decays to a highly visible final state, and that reconstruction
efficiencies are perfect.  One can then use the standard QCD diagrams to
calculate the largest value of $m_Q$ for which the Tevatron would observe 10
$Q\ov Q$ pair production events with 10~fb$^{-1}$ of data.  In this idealized
example, the hypothetical Tevatron bound is $m_Q \gtrsim 500 \GeV$.  The same
exercise can be repeated for the LHC as a function of the center of mass energy
and integrated luminosity, and the result is shown in Fig.~\ref{QCDpairs}.  

To reach the Tevatron sensitivity for QCD pair production at a 7 TeV LHC, the
required luminosity is about 50~pb$^{-1}$.  While this is likely within the
reach of an early LHC run, the LHC will not easily surpass Tevatron bounds in
this channel, and it is unlikely that a 5$\sigma$ LHC discovery is possible
without the Tevatron already having seen some events.  (The same holds for
colored scalar pair
production~\cite{Manohar:2006ga,Dobrescu:2007yp,Burgess:2009wm}.)  This
conclusion is only bolstered when realistic branching fractions to visible final
states and signal efficiencies are taken into account.

The primary reason why QCD pair production is not a supermodel is that the same
final state can also be produced from the $q\bar q$ initial state, where the LHC has
less of an advantage over the Tevatron.  The
situation can be improved if there is a large multiplicity of near-degenerate
new colored states or if the new states are color octets (like gluinos in
supersymmetry).  Then the total cross sections are larger by a multiplicity
factor and the LHC reach can surpass that of the Tevatron (where the cross
section is more strongly suppressed at higher masses).  As an example,
leptoquark pair production~\cite{Leptoquark_pdg} yields the easily reconstructable final state of two
leptons and two jets, so this could be a supermodel  with a sufficiently large
multiplicity of such leptoquarks.

In any case, because QCD pair production is
quite well-studied in specific new physics scenarios and because the early LHC
advantage over the Tevatron can only be marginal, we will not consider it to be
a supermodel in this paper.  In Sec.~\ref{subsec:Resurect_Pair_Prod}, though,
we show that pair production through an intermediate resonance can give
rise to supermodels.

\subsection{Resonance production}

While pair production of new colored particles is not a supermodel, production
of an $s$-channel resonance has the potential to be a supermodel, as long as the
resonance has renormalizable couplings to the partonic initial states.  Recall
that parametrically the production cross section for a single resonance is
enhanced over pair production by a phase space factor of $16 \pi^2$.  Moreover,
unlike QCD pair production where $SU(3)$ gauge invariance relates the $gg$ and
$q\bar{q}$ scattering amplitudes, single resonance production can be dominated
by one partonic initial state.

In the narrow width approximation, we can parametrize generic single resonance
production by
\be
\sigma(p_i p_j \to X) = [g^2_{\rm eff}]_{ij}\, \delta(\hat{s} - m_X^2) \,,
\ee
where $p_{i,j}$ denote the two partons which participate in the hard scattering,
$m_X$ is the mass of the resonance, and $[g^2_{\rm eff}]_{ij}$ encodes all
information about the production of resonance $X$ from the two partons,
including couplings, polarization, and color factors.  Using the parton
luminosities defined in  Eq.~\Eq{eq:partonluminosity}, the hadronic cross
section is
\begin{eqnarray}
\sigma(p p \to X)
&=& \frac{1}{m_X^2}\sum_{ij}\, [g^2_{\rm eff}]_{ij}\, {\cal F}_{ij}(m_X^2, s)\,.
\end{eqnarray}
For the resonances considered in this paper, one production channel dominates,
allowing us to drop the $ij$ label from $g^2_{\rm eff}$.

For reasonably narrow resonances with dimension four couplings, $g^2_{\rm eff}$
can be order $1$, which is the case for the $q\bar{q}$ and $qq$ initial
states.  However, for the $qg$ or $gg$ initial states $SU(3)$ gauge invariance
forbids renormalizable couplings to a single resonance.  For example, for the
$gg$ initial state, the lowest dimension operator allowed is a dimension five
coupling to a scalar or pseudoscalar:
\be
\label{eq:glueglueX}
\frac{g_s^2}{16 \pi^2 \Lambda}\, X\, \Tr(G_{\mu\nu} G^{\mu\nu}) \,.
\ee
The coefficient of the operator has been estimated assuming perturbative physics at $\Lambda \sim$ TeV, with the $1/(16 \pi^2)$ factor coming from a loop. This gives a rather suppressed effective coupling $[g^2_{\rm eff}]_{gg} \sim [1/(16 \pi^2) (m_X/\Lambda)]^2 \sim [1/(16 \pi^2)]^2$, where we have assumed $m_X$ is around a TeV.  If there is strong dynamics involving $X$ at the TeV scale, then the coefficient in Eq.~\Eq{eq:glueglueX} can be enhanced up to its naive dimensional analysis value $g_s^2/(4 \pi \Lambda)$. However, such strong dynamics near the TeV scale is constrained by precision measurements, and we will adopt the perturbative estimate $g^2_{\rm eff} \sim [1/(16 \pi^2)]^2$ for both $gg$ and $qg$ resonances.


\begin{figure*}[bt]
\includegraphics[width=0.45\textwidth]{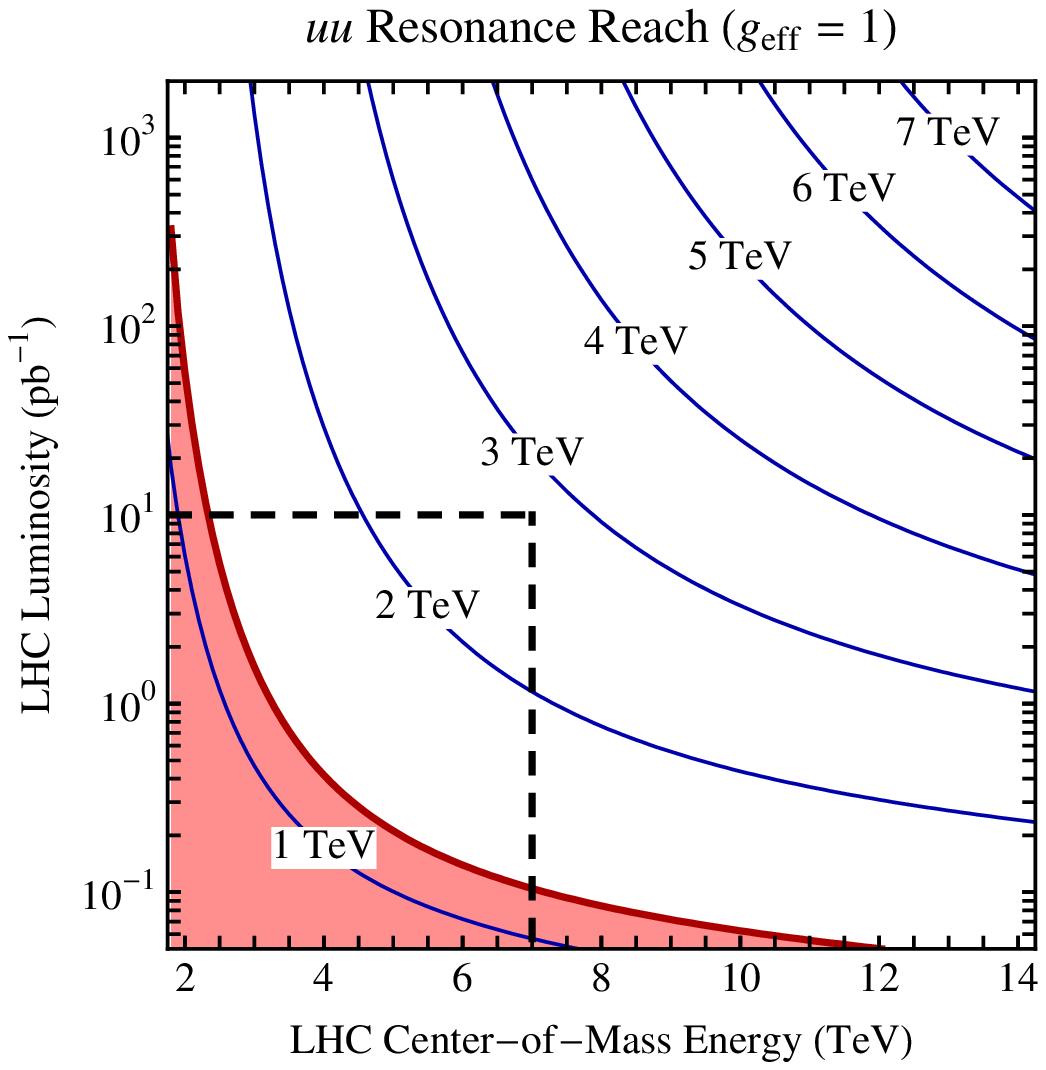}
\hfil
\includegraphics[width=0.45\textwidth]{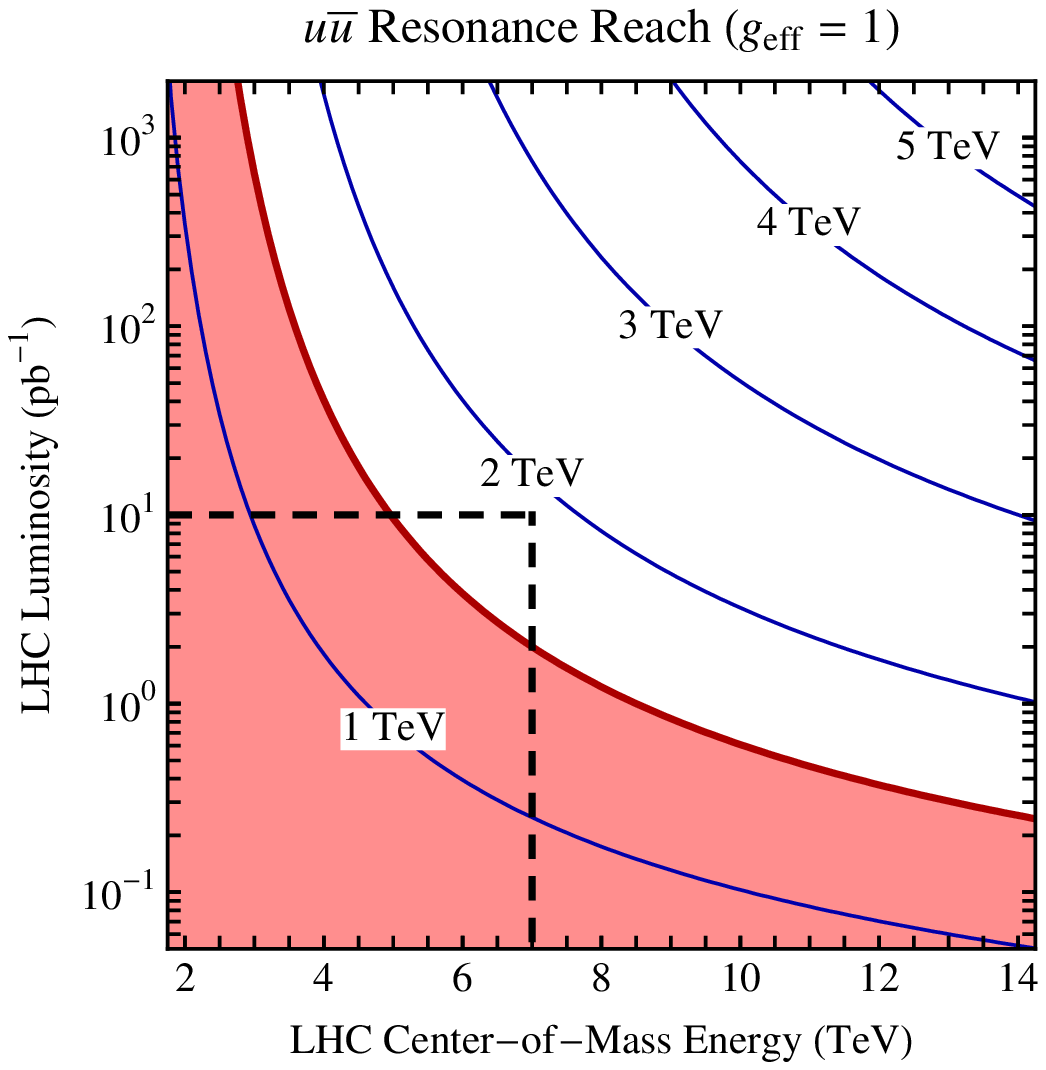}

\vspace{10pt}
\includegraphics[width=0.45\textwidth]{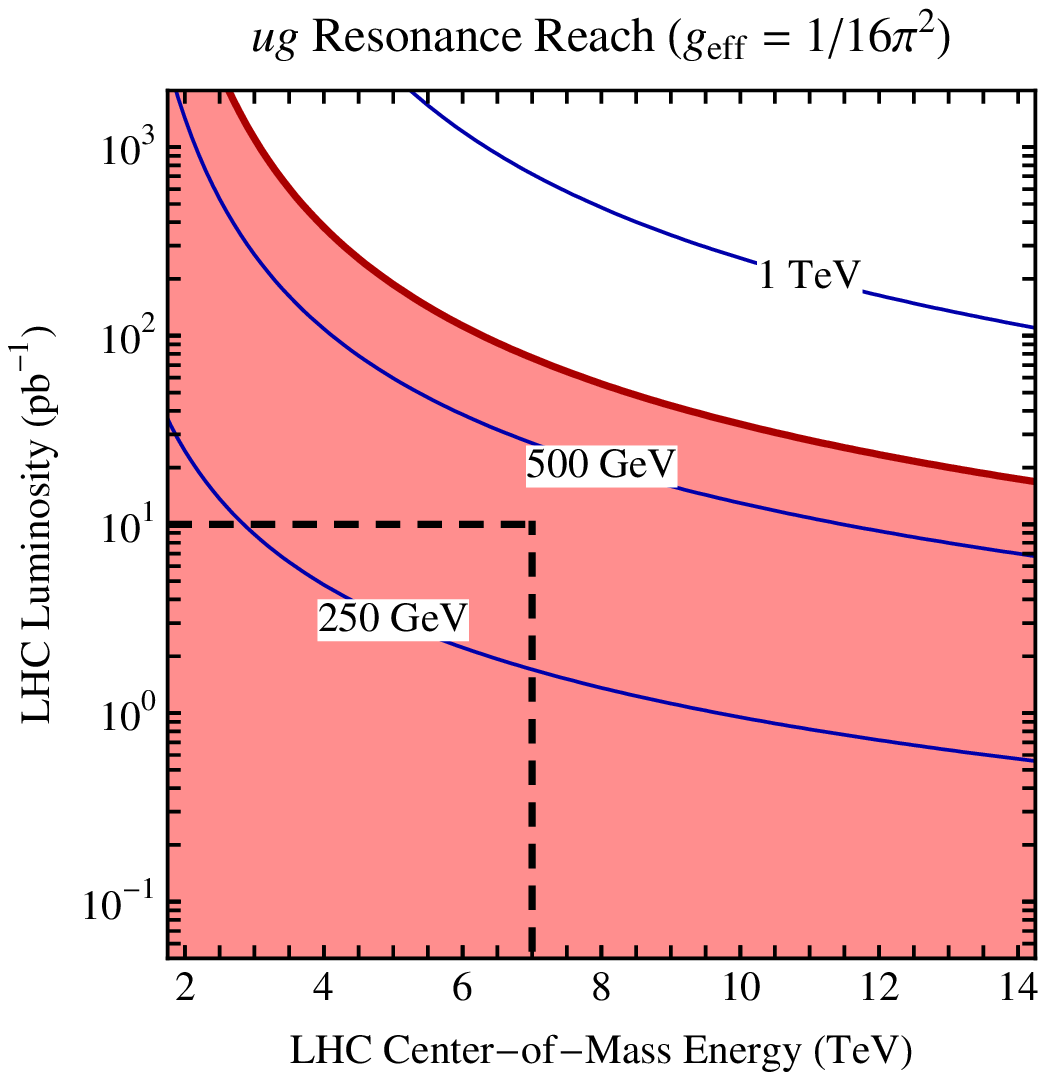}
\hfil
\includegraphics[width=0.45\textwidth]{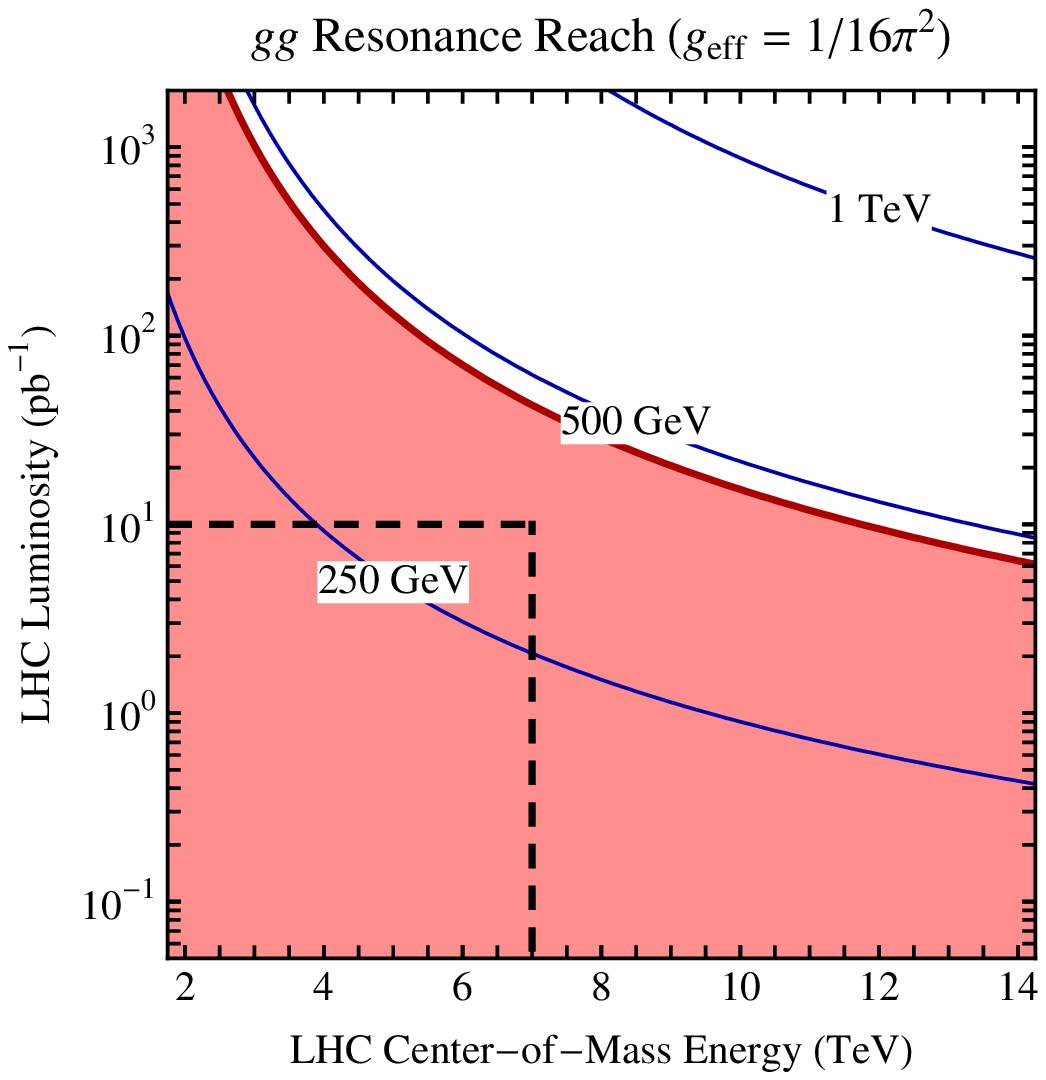}
\caption{LHC reach for single resonance production as a function of energy and
luminosity.  As in Fig.~\ref{QCDpairs}, the contours show the production of 10
events for a given resonance mass, the red regions show the Tevatron sensitivity
with 10~fb$^{-1}$, and the intersection of the dashed lines shows the maximum
resonance mass which can be probed by the 7 TeV LHC with 10~pb$^{-1}$ data.
The expected couplings for perturbative new physics in
Eq.~(\ref{Nijpert}) are included.  One sees that the early LHC can exceed the
Tevatron sensitivity for $q\bar q$ and especially for $qq$ resonances.}
\label{fig:foursimpleplots}
\end{figure*}

\begin{figure*}[t]
\includegraphics[width=0.45\textwidth]{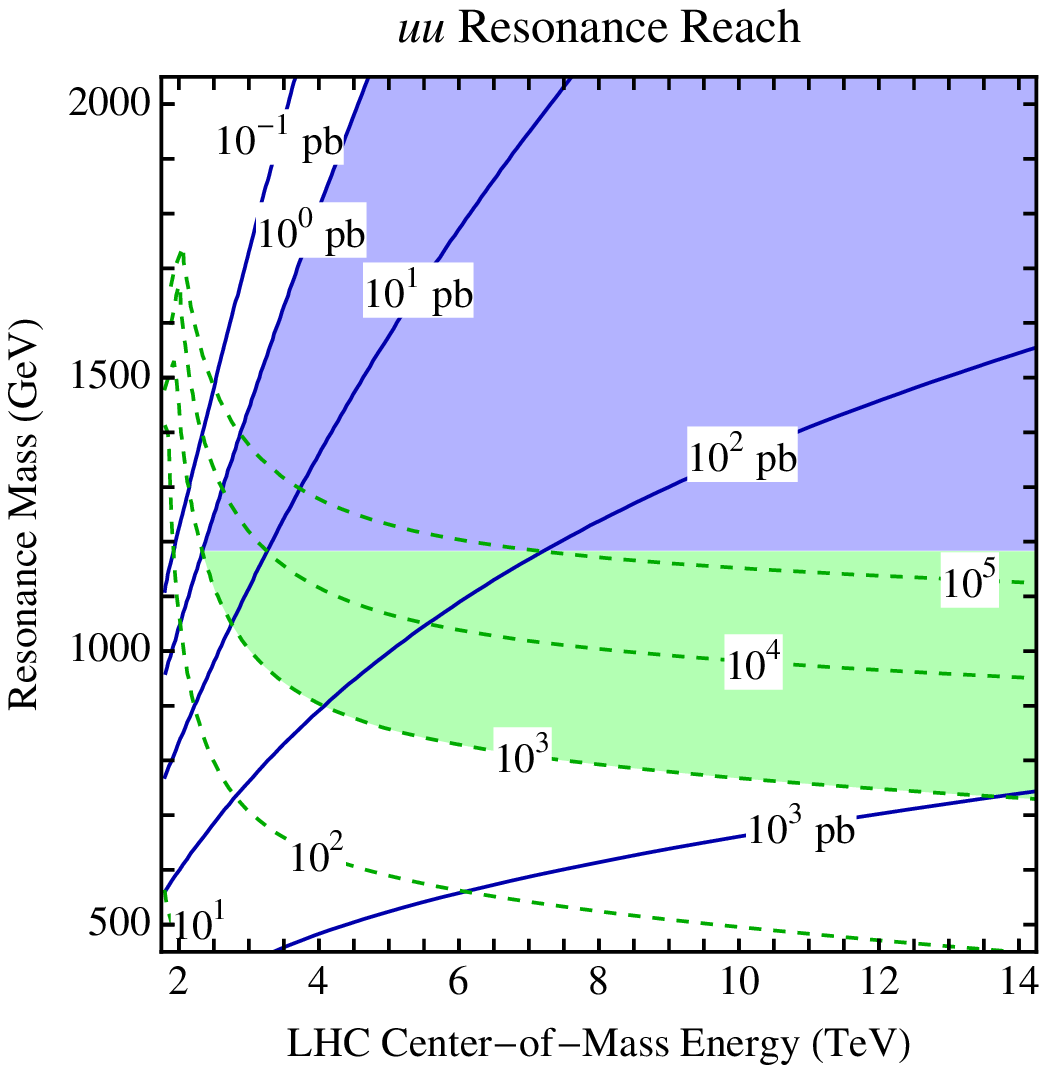}
\hfil
\includegraphics[width=0.45\textwidth]{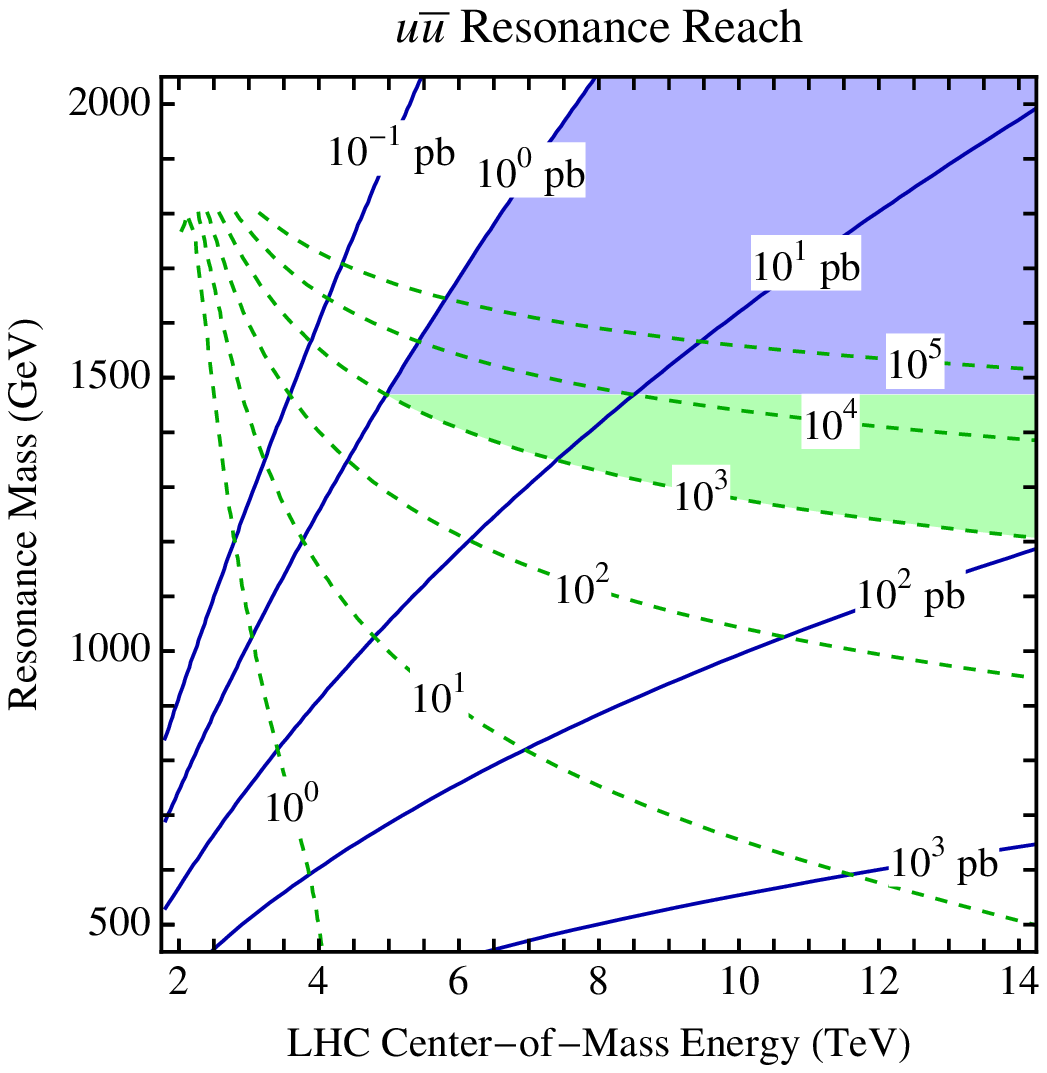}
\caption{The LHC reach for $uu$ and $u\bar u$ resonances in the LHC energy vs.\
resonance mass plane.  The solid lines are contours of constant LHC production
cross sections for $g^2_{\rm eff} = 1$, and the dashed green lines are contours
of constant LHC to Tevatron cross section ratios.  The blue shaded regions show
where the discovery reach of a 10~pb$^{-1}$ LHC run is beyond that of the 10~fb$^{-1}$ Tevatron. 
The green regions show where the LHC sensitivity is greater than that of the
Tevatron, but the Tevatron can also see at least 10 events.}
\label{fig:twowedgeplots}
\end{figure*}

In Fig.~\ref{fig:foursimpleplots}, we show our estimate of the generic early LHC reach
in $m_X$, as a function of the energy and luminosity, for the
four resonance channels using
\be\label{Nijpert}
g^2_{\rm eff} = \cases{ 1 \,, &  $qq$ or $q\bar{q}$ resonances\,; \cr
  \big[1/(16\pi^2)\big]^2 \,,  &  $qg$ or $gg$ resonances\,. \cr}
\ee
As in Fig.~\ref{QCDpairs}, we assume 100\% branching fraction of $X$ to highly
visible final states and assume perfect detector efficiency, though we will
relax these assumptions below.  Note that for a charged resonance (that is
produced from $qq$ and $qg$ initial states), the search strategy for the charge
conjugate resonance arising from $\bar q \bar q$ and $\bar q g$ initial states
is equivalent, and any analysis will include both.  The plots in
Fig.~\ref{fig:foursimpleplots} use the $u$ quark parton distribution function
for simplicity (rather than that of the $d$ or both) and include the $uu + \bar
u \bar u$ and $ug + \bar u g$ initial states instead of $uu$ and $ug$ only.

We see that while gluons are by far the most abundant partons at small $x$,
scattering initiated by gluons does not lead to very large resonance cross
sections at the LHC, if perturbative couplings are assumed.  In the $u\bar u$
and especially in the $uu$ channels, shown in the two upper plots of
Fig.~\ref{fig:foursimpleplots}, the first LHC
run even with modest energy and luminosity will supersede the Tevatron.
Thus, $qq$ and $q\bar q$ resonances are the most suitable starting
points for constructing supermodels, examples of which will appear in Sec.~\ref{sec:decay}.

\subsection{\boldmath Production of $qq$ and $q\ov q$ resonances}

The plots in Fig.~\ref{fig:foursimpleplots} give a rough idea of the LHC
discovery potential for $s$-channel resonances.  They are valid for a
particular value of the effective coupling, $g^2_{\rm eff}$, and assume that the
$X$ resonance is observed with 100\% efficiency.  For the two most promising
scenarios of $q q$ and $q\ov q$ resonances, we are interested in the dependence
of the reach on $g^2_{\rm eff}$ and on branching fractions/efficiencies. Here,
we introduce a new kind of plot which is convenient for reading off cross
sections at the LHC and comparing them to the Tevatron for variable couplings
and detection efficiencies.
In Fig.~\ref{fig:twowedgeplots}, we plot in the LHC energy vs.\ resonance mass
plane the contours of constant production cross section and contours of constant
ratio of LHC vs.\ Tevatron cross section. 

The solid curves (with positive slopes) in Fig.~\ref{fig:twowedgeplots} show
contours of constant LHC cross sections for $g^2_{\rm
eff}=1$.
From these, one can read off how many events are produced for a given LHC
luminosity as a function of the resonance mass and the LHC energy.
For example, assuming 100\% visible decay rate and detection efficiency, the
region to the right of the curve labeled ``$10^0$~pb'' will yield at least 10
events with $10~\pb^{-1}$ of LHC data.  For a concrete model, with a different
value of $g^2_{\rm eff}$, branching ratio $\Br$ to easily reconstructable final
state(s), and detector/reconstruction efficiency $\mathrm{Eff}_{\rm LHC}$, the
LHC with a given energy and $\mathcal{L}_{\rm LHC}$ luminosity can see $N$ or
more events to the right of the solid curve labeled
\be
\label{eq:solidcurveinfo}
\frac{N}{\mathcal{L}_{\rm LHC} \times g^2_{\rm eff}
  \times \Br \times \mathrm{Eff}_{\rm LHC}} \,.
\ee
For example, if $g^2_{\rm eff} = 0.1$ and $\Br \times \mathrm{Eff}_{\rm LHC} =
10\%$, then 10 or more events can be observed in the region to the right of the
curve labeled ``$10^2$~pb" (``$10^1$~pb") with 10~pb$^{-1}$ (100~pb$^{-1}$) of
LHC data.

The dashed curves (with negative slopes) in Fig.~\ref{fig:twowedgeplots} show
contours of constant ratio of LHC vs.\ Tevatron cross sections.
From these, one can read off the relative advantage of the LHC compared to the
Tevatron for a given model.  For many resonances, the same initial state
dominates the production of a resonance at the Tevatron and the LHC, so that the
factors of $g^2_{\rm eff}$ cancel in the ratio of the two cross section. If, in
addition, the same final states are searched for at the Tevatron and at the LHC
and the detection efficiencies are similar, then these cancel in the ratio as
well.  Therefore, the LHC has better sensitivity than the Tevatron in the region
above the dashed curve labelled by
\be
\mathcal{L}_{\rm TEV} / \mathcal{L}_{\rm LHC} \,.
\label{eq:ratios}
\ee
For example, the LHC with $10~\pb^{-1}$ (100~pb$^{-1}$) will produce more
events than the Tevatron with $10~\fb^{-1}$ in the region above the dashed curve
labeled ``$10^3$" (``$10^2$"). Any significant differences in detection 
efficiencies between the two colliders can easily be included by multiplying
Eq.~(\ref{eq:ratios}) by $\mathrm{Eff}_{\rm TEV} / \mathrm{Eff}_{\rm LHC}$.

Thus, it is a ``wedge" bounded by a solid and a dashed curve which defines the
region in which the LHC has better sensitivity than the Tevatron and yields at
least a certain number of events.  For example, the wedge to the right of the
intersection of the ``$10^0$~pb'' and the ``$10^3$'' curves gives the region for
which at least 10 events are produced with $10~\pb^{-1}$ of LHC data and the
number of events at the LHC is greater than that at the Tevatron.  These regions
are shaded in Fig.~\ref{fig:twowedgeplots}.  (To include model specific effects,
replace the ``$10^0$~pb'' solid curve by the $10^0 {\rm pb}/(g^2_{\rm eff}\,
\Br\, \mathrm{Eff}_{\rm LHC})$ one.)

At the intersection of a solid and a dashed curve, the ratio of their labels
gives the Tevatron cross section, and can be used to estimate the Tevatron
discovery reach.  
The intersection of any ``$10^{n+a}$~pb" solid curve with a
``$10^{a}$" dashed curve corresponds to the same fixed Tevatron cross section
of $10^n$~pb for arbitrary $a$. Since the Tevatron cross section does not
depend on the LHC energy, these intersections lie on a horizontal
line.  The corresponding value of the resonance mass is the one for which the
Tevatron with 10~fb$^{-1}$ data produces $10^{4+n}$ events.  For example, for
masses below the straight line across the intersection of the ``$10^0$~pb''
and the ``$10^3$'' curves (i.e.\ $n = -3$), the Tevatron will also produce at least 10 events
with 10~fb$^{-1}$ data.  While everywhere in the full wedges shaded in
Fig.~\ref{fig:twowedgeplots} the LHC sensitivity is better than that of the
Tevatron, below this straight line the Tevatron sensitivity is good enough that
it can make a discovery prior to the LHC.\footnote{For a sequential $Z'$
decaying to leptons, taking $g^2_{\rm eff}\, \Br\, \mathrm{Eff} \sim 0.01$, this
predicts a 10~fb$^{-1}$ Tevatron bound near 1.2~TeV; crude, but
reasonable~\cite{Aaltonen:2008ah}.}  Therefore, it is only the part of the
``wedge" above this straight line, shaded blue, which is the true LHC discovery
region.

Using these plots, one can also estimate the minimum value of $m_X$ and
$g^2_{\rm eff}\, \Br\, \mathrm{Eff}_{\rm LHC}$ for a scenario to be a
supermodel.  Take a $q \bar q$ resonance as an example.  A 7~TeV and
$10$~pb$^{-1}$ early LHC run supersedes the Tevatron sensitivity for a $m_X
\gtrsim 1.4$~TeV (the value of the ``$10^3$" dashed curve at 7~TeV).  We can
then read off that $g^2_{\rm eff}\, \Br\, \mathrm{Eff}_{\rm LHC} \gtrsim 0.1$ is
required to observe at least 10 events.  For larger initial LHC luminosity, this
minimum of course gets smaller.

\begin{figure*}[tb]
\centerline{\includegraphics[height=3cm]{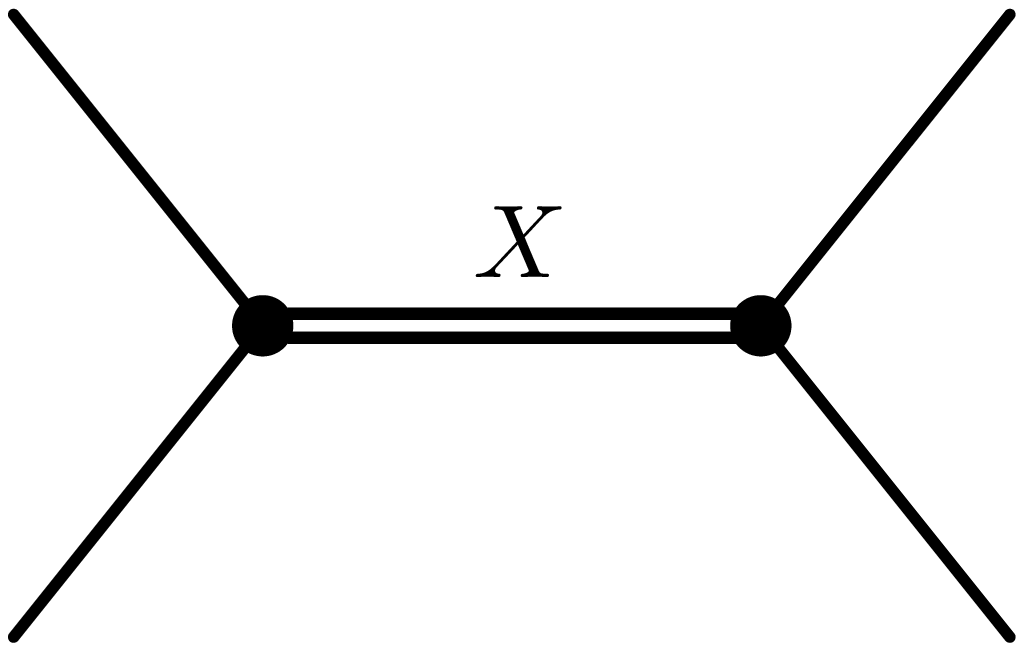}\hfil
  \includegraphics[height=3cm]{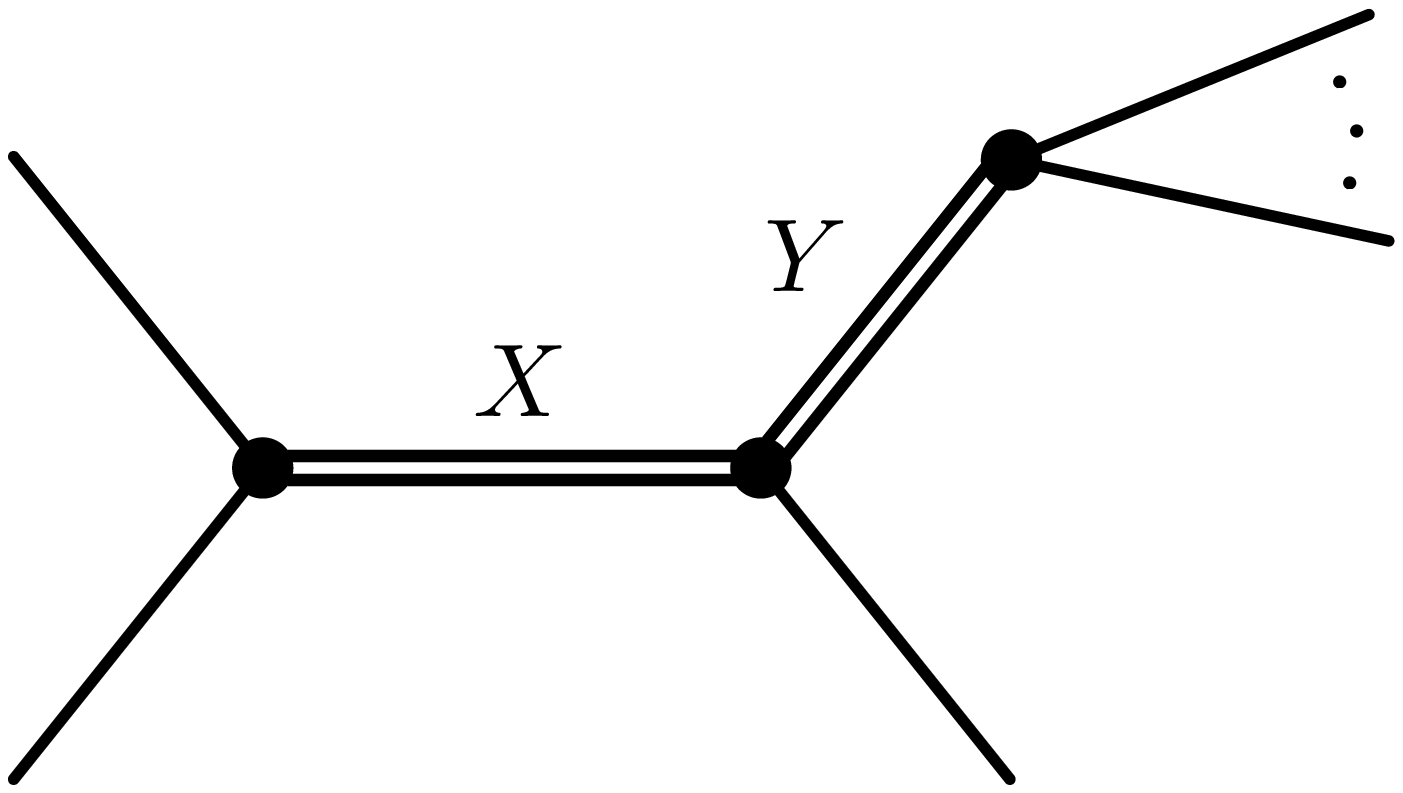}\hfil
  \includegraphics[height=3cm]{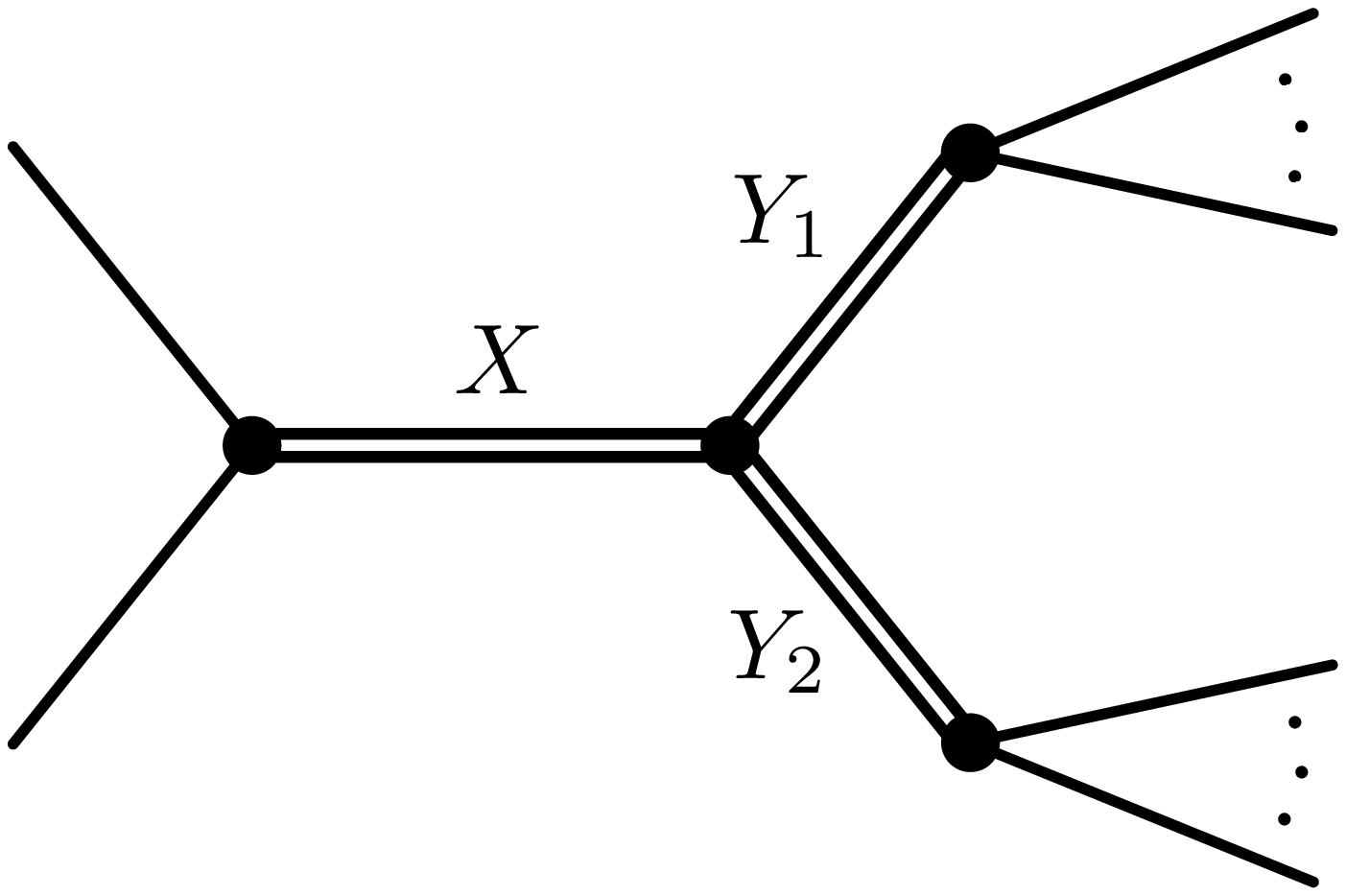}}
\centerline{\footnotesize (A) \hspace*{4.5cm} (B) \hspace*{5cm} (C)}
\caption{Three typical event topologies for weakly coupled resonance decay. 
Double lines denote new particles, while single lines are detectable final states,
either standard model particles or new (quasi-)stable states.
The ellipses indicate that the secondary resonances may
decay to two or more particles.}
\label{fig:topologies}
\end{figure*}

\section{Example Supermodels}
\label{sec:decay}

Considering production cross sections alone, $q\bar{q}$ and $qq$ resonances
emerged as the two best starting points for constructing early LHC supermodels. 
In this section, we consider concrete supermodel examples to see what kind of
final states can be obtained from the decay of these resonances.  Since we are
interested in final states that involve the cleanest signatures and least
background contamination, we concentrate on decay chains yielding (at least) two
charged leptons (or two other stable charged particles) in the final state. 

Because there is a plethora of possible decay patterns for a $qq$ or $q\bar{q}$
resonance, we find it convenient to classify the decay modes of new resonances
in terms of three basic decay topologies, which often appear in perturbative new
physics scenarios.  These are depicted in Fig.~\ref{fig:topologies}.  Double
lines denote new resonances and solid lines are detectable final states, either
standard model particles or new (quasi-)stable states.
\begin{itemize}
\item Topology A:  The resonance $X$ decays directly to two detectable final
state particles.  Note that a three-body decay for the resonance will
generically not compete with the decay back to QCD partons, since the branching
fraction to a three-body final state is suppressed by a phase space factor
compared to the two-body decay channel back to the initial state.
\item Topology B:  The resonance $X$ decays to one detectable final state and
one new secondary resonance $Y$.  Subsequently $Y$ can decay to two or more
standard model particles.  Note that $Y$ always has a decay mode back through
the production diagram with a virtual $X$.
\item Topology C:  The resonance $X$ decays to two new secondary resonances
$Y_1$ and $Y_2$, possibly of different masses.  These secondary resonances can
each decay to two or more detectable states.  This topology can be used
for example to ``resurrect'' pair production of colored particles as a
supermodel.
\end{itemize}

The classic case of topology A is a leptonically decaying $Z'$, and this is
often described as the simplest new physics channel to discover at a hadron
collider.  However, we will argue below that a standard $Z'$ is not a supermodel
if one takes into account the indirect constraints coming from LEP II.  After
dismissing the standard $Z'$ case, we will give a (non-exhaustive) list of
supermodels exhibiting the three above topologies.  

While some of the example supermodels are simple variants on well-motivated
models, the case of a $qq$ resonance (or diquark) is less well-known (though we
will show that such a particle can arise in the
minimal supersymmmetric standard model; 
see also Ref.~\cite{Angelopoulos:1986uq}).  Despite
the unfamiliar quantum numbers of the diquark, the final states achievable in
diquark decays can be shared by more familiar resonances, albeit with smaller
non-supermodel cross sections.  In this way, searches for supermodels in early
LHC data can anticipate searches that require higher \mbox{luminosity}.  

For reference, the standard model quantum numbers are:
\be\label{charges}
\begin{tabular}{c|c|c|c|r}
Field & Spin & $SU(3)_C$ & $SU(2)_L$ & $U(1)_Y$ \\ 
\hline
$q$ & $1/2$ & $\mathbf{3}$ & $\mathbf{2}$ & $1/6$ \\
$u^c$ & $1/2$ & $\mathbf{\bar{3}}$ & $-$ & $-2/3$ \\
$d^c$ & $1/2$ & $\mathbf{\bar{3}}$ & $-$ & $1/3$ \\
$l$ & $1/2$ & $-$ & $\mathbf{2}$ & $-1/2$ \\
$e^c$ & $1/2$ & $-$ & $-$ & $1\phantom{/2}$ \\
\hline
$h$ & $0$ & $-$ & $\mathbf{2}$ & $1/2$
\end{tabular}
\ee

\subsection{\boldmath The case against a standard $Z'$}

For a $q\bar{q}$ resonances to be supermodel, it must have a large branching
fraction to visible final states.  In particular, since a $q\bar{q}$ resonance
can have zero electric charge, it is natural for such a resonance to decay via
topology~A to pairs of oppositely charged leptons, in particular $e^+e^-$ and
$\mu^+\mu^-$.  However, the same resonance also induces a low energy effective
four-lepton vertex, and such operators are severely constrained by LEP~II.  As
recently emphasized in Ref.~\cite{Salvioni:2009mt}, once the LEP~II bound is imposed,
the branching fraction of the $q\bar{q}$ resonance to $\ell^+\ell^-$ has to be
too small to realize a supermodel.

As a concrete example, consider a new $U(1)$ gauge boson $Z'$:
\be
\begin{tabular}{c|c|c|c|c}
Field & Spin & $SU(3)_C$ & $SU(2)_L$ & $U(1)_Y$ \\ 
\hline
$Z'$ & $1$ & $-$ & $-$ & $-$
\end{tabular}
\ee
with couplings to standard model fermions $\psi_i$ of the form\footnote{Here and
below, we use 2-component fermion notation.}
\be\label{OZ}
{\cal O}_{Z'} = g_i\, Z'_\mu\, \overline{\psi_i}\, \bar{\sigma}^\mu \psi_i \,,
\ee
where $g_i$ are the corresponding coupling constants.  For simplicity, imagine a
common coupling $g_Q$ to quarks and $g_L$ to leptons.\footnote{While such a choice of
$Z'$ charges is typically anomalous (unless $g_Q = - 3 g_L$), such anomalies can
always be canceled with new fermions (spectators) whose masses result from the
same $U(1)$ symmetry breaking which gives the $Z'$ its mass.}

In the narrow width approximation, the production cross section for this $Z'$
resonance is 
\be
\sigma(q\bar q \to Z') = \frac\pi{3}\, g_{Q}^2 \, \delta(\hat s - m_{Z'}^2)\, ,
\ee
and the branching ratio of this $Z'$ to charged leptons is\footnote{We
are assuming Dirac neutrino masses for definiteness, and we do not consider
final state $\tau$'s to be easily reconstructable, hence the overall factor of $2/3$.}
\be
\Br(Z' \to \ell^+ \ell^-) = \frac{2}{3} \frac{g_L^2}{2 g_L^2 + 6 g_Q^2} \,.
\ee
The coupling to $Z'$ to leptons is strongly constrained by LEP II limits on
four-fermion operators~\cite{Alcaraz:2006mx},
\be
\frac{g_L^2}{m_{Z'}^2} \le \frac{4\pi}{(10 \TeV)^2} .
\ee
Putting this together, we find
\be
g^2_{\rm eff} \times \Br = \frac{2\pi}{9}\, \frac{g_Q^2\, g_L^2}{2 g_L^2 + 6 g_Q^2}
  \lesssim \left(\frac{m_{Z'}}{8 \TeV}\right)^2 \,. 
\ee
Using Fig.~\ref{fig:twowedgeplots} and Eq.~\Eq{eq:solidcurveinfo}, one finds that there is no value of $m_{Z'}$
for which 10 $Z' \to \ell^+ \ell^-$ events could be seen with $10~\pb^{-1}$ of
data, even with a center of mass energy of 14~TeV.

There are ways to evade this conclusion.  Since the LEP~II bound only
applies for the electron coupling, one could imagine coupling the $Z'$
only to muons.  However, such flavor non-universal couplings typically require
significant fine-tuning to avoid constraints from flavor changing neutral currents.
Alternatively, one might
consider the production of a $W'$-like resonance, i.e., a resonance with
electric charge 1, which decays to a charged lepton and a neutrino.  However,
typically such $W'$ models have an accompanying $Z'$-like state, which faces the
strong LEP~II bounds.

\subsection{Decays to quasi-stable particles}
\label{sec:decaytostable}

While the decay of a $Z'$ to standard model charged leptons does not give a
viable supermodel example of topology A, one could imagine a $q\bar{q}$
resonance that instead decayed with a large branching fraction to new
quasi-stable charged particles.  Since ATLAS and CMS trigger on penetrating
charged particles as if they were muons~\cite{Fairbairn:2006gg}, such
scenarios are as visible  in the early LHC data as a $Z'$ decaying to muons.  

One simple choice is to take the above $Z'$ boson with $g_L = 0$, and introduce
$N_E$ vector-like ``leptons'', $E$ and $E^c$, with coupling $g_E$ to the $Z'$
and non-zero electric charge.
\be
\label{eq:stableleptonqnumbers}
\begin{tabular}{c|c|c|c|c}
Field & Spin & $SU(3)_C$ & $SU(2)_L$ & $U(1)_Y$ \\ 
\hline
$E$ & $1/2$ & $-$ & $-$ & $-1$ \\
$E^c$ & $1/2$ & $-$ & $-$ & $\phantom{-} 1$ \\
\end{tabular}
\ee
The $E$ and $E^c$ fields can have an approximate $\mathbf{Z}_2$ symmetry to make
them long-lived, but they must eventually decay to avoid cosmological bounds on
absolutely stable charged relics.  For this reason, the hypercharge for $E$ and
$E^c$ has been chosen such that the quasi-stable lepton can decay via
mixing with the standard model $e^c$.

The branching fraction of the $Z'$ to the stable charged states is
\be
\Br(Z' \to E^+ E^-) = \frac{N_E\, g_E^2}{N_E\, g_E^2 + 18 g_Q^2} \, .
\ee
For large enough values of $N_E\, g_E^2$, the branching fraction can be order
1.  Because the stable ``leptons'' are being produced from a resonance, they
will typically have a velocity of
\be
\beta  \simeq \sqrt{1 - \frac{4m_E^2}{m_{Z'}^2}} \, ,
\ee
so depending on the relevant mass ratio $m_E/ m_{Z'}$, a standard ``slow muon''
cut of $\beta < 0.9$ \cite{CMS_stable_charged}  will not be particularly
effective at capturing the signal.  In such cases, these stable charged
particles should be treated like ordinary muons (without imposing any kind of
$dE/dx$ quality cut tailored to muons) to reconstruct a $Z'$ resonance.

Alternatively, one could consider a $Z'$ that decayed to quasi-stable colored
particles that then form $R$-hadron-like bound states with QCD partons.  Such
$R$-hadron final states could potentially be visible in early LHC data, though
charge flipping interactions \cite{Fairbairn:2006gg} complicate both triggering
and momentum reconstruction.

\subsection{Fun with diquarks}

From Fig.~\ref{fig:twowedgeplots}, one sees that $qq$ resonances can yield an
impressive early LHC reach.  Such resonances are known as diquarks, and they
have spin zero or one, carry baryon number 2/3, and electric charge 4/3, 1/3 or
$-$2/3. They may transform as a $\bf 6$ or $\bf \overline 3$ of color. Their
couplings are necessarily non-trivial in flavor space  because the initial
quarks carry flavor.  Flavor changing neutral currents impose constraints on
couplings of new states with masses of order TeV and large couplings to first
generation quarks.  To be safe, we consider diquarks with the same flavor
quantum numbers as the quarks which produce them, allowing the couplings of the
diquark to quarks to be flavor invariant.  (Other models~\cite{Chacko:1998td}
consider a single diquark which couples to the quarks with a $3\times3$ Yukawa
matrix, and generically lead to dangerous flavor
violation.)

To be concrete, we consider a spin-zero and color-six diquark, with couplings to
the $SU(2)$ singlet up-type quarks only and symmetric in flavor
indices.\footnote{We discuss the phenomenology of a color-triplet diquark which
occurs naturally in the context of R-parity violating SUSY in the next
subsection.} 
\be
\begin{tabular}{c|c|c|c|c}
Field & Spin & $SU(3)_C$ & $SU(2)_L$ & $U(1)_Y$ \\ 
\hline
$D$ & $0$ & $\mathbf{6}$ & $-$ & $4/3$\\
\end{tabular}
\ee
The production operator can be written as
\be\label{OD}
{\cal O}_D = \frac{\kappa_D}{2}\, D\, u^c\, u^c \,,
\ee
where $u^c$ are the up-type singlet quarks and $D$ is the diquark.  The
normalization of the coupling constant $\kappa_D$ depends on the normalization
of the color matrices $R^a$ with which we expand $D=D^a R^a$. We use orthonormal
matrices such that ${\rm Tr}(R^a R^b) = \delta^{ab}$.  Then the partonic cross
section is\footnote{Recently radiative corrections were calculated, giving a
$K$-factor of 1.3~\cite{Mohapatra:2007af}.}
\be\label{sigmaD}
\sigma(uu \to D) =  \frac{\pi}{6}\, \kappa_D^2\, \delta(\hat s-m_D^2) \,.
\ee

If Eq.~\Eq{OD} were the only available coupling for the diquark, then any
produced diquark would simply decay back to the initial state with a partial
width given by $\Gamma = \kappa_D^2 m_D / (16\pi)$.  To be a supermodel, the
diquark has to have a large branching fraction to a visible final state.  By
color conservation, diquark decays must yield at least two jets in the final
state, so the most $Z'$-like decay possible for a diquark (in the sense that it
yields two oppositely charged leptons in the final state) is
\begin{equation}
\label{eq:desiredDdecay}
D \to 2j + \ell^+ \ell^-.
\end{equation}
Such a final state can appear via topology B or C, though we will consider the
case of topology B since it requires fewer new degrees of freedom.

For example, we can introduce a vector-like fermion $L$ and $L^c$, with the
quantum numbers:
\be
\label{eq:leptodiqnumbers2}
\begin{tabular}{c|c|c|c|c}
Field & Spin & $SU(3)_C$ & $SU(2)_L$ & $U(1)_Y$ \\ 
\hline
$L$ & $1/2$ & $\mathbf{6}$ & $-$ & $\phantom{-} 7/3$ \\
$L^c$ & $1/2$ & $\mathbf{\overline{6}}$ & $-$ & $-7/3$ \\
\end{tabular}
\ee
Given its quantum numbers, $L$/$L^c$ would be called a ``leptodiquark''.  The
diquark can then decay via the operator
\be
\label{eq:leptodiquarkcouplings}
\bar{\kappa}_D\, D\, L^c\, e^c
\ee
with a decay width of $\Gamma = \bar\kappa_D^2 m_D/(16 \pi)$.
Thus, as long as $\bar \kappa_D \gtrsim \kappa_D$, the diquark
preferentially decays to the leptodiquark and a lepton.  The $L^c$ will finally
decay via the operators in \Eqs{OD}{eq:leptodiquarkcouplings} as
\be
L^c \to \ov{e^c}\, u^c u^c
\ee
through an off-shell diquark, leading to the full decay chain:
\be
\begin{array}{l}
uu \rightarrow D \\[-2pt]
\phantom{uu \rightarrow \ \, } \bentarrow \ell^- L \\[-2pt]
\phantom{uu \rightarrow \ \,  \bentarrow \ell^-} \bentarrow \ell^+ \, 2j \,.
\end{array}
\ee
The final state therefore has two jets plus
an opposite sign lepton pair, arranged in topology B.

While this diquark-leptodiquark system may strike the reader as baroque, 
the identical decay topology appears in the case of a $W_R'$
gauge boson~\cite{Aad:2009wy,Ball:2007zza}, where the diquark plays the role of
the $W_R'$ and the leptodiquark plays the role of a right-handed neutrino. 
However, discovering a left-right symmetric model through this channel 
typically requires $1~\fb^{-1}$~\cite{Aad:2009wy,Ball:2007zza} of LHC data, whereas
 the diquark-leptodiquark example motivates a search for the $2j + \ell^+
\ell^-$ final state in early LHC data.

\subsection{Squarks as diquarks}

A curious example of a supermodel with diquarks is the MSSM with $R$-parity
violation. The superpotential term $\lambda_{ijk} U_i^c D_j^c D_k^c$, where
$U^c$ and $D^c$ are chiral superfields, allows for resonant production of a
single squark in the scattering of two quarks \cite{Dimopoulos:1988fr}.  The
coupling constants $\lambda_{ijk}$ are constrained by flavor physics,
$N-\overline N$ oscillations, and double-nucleon decays. However for squark
masses of order 1 TeV or larger, couplings of order one for $\lambda_{112}$ and
$\lambda_{113}$ are allowed \cite{Goity:1994dq,Allanach:1999ic}.

Such order one couplings give a very large cross section for single squark
production in $ud$ scattering. Contours of these cross sections are similar
to what is shown in the left panel of Fig.~\ref{fig:twowedgeplots} for $uu$
scattering.  The produced squarks ($\tilde s^c$ or $\tilde b^c$) have order one
branching fractions to decay back into dijets by the inverse of the production
process. More interestingly, they also have large branching fractions to decay
via one of the typical SUSY cascade decays ending in the lightest superpartner
(LSP).  The decay chains depend on the details of the superpartner spectrum. 

For example, if the order of superpartner masses from heaviest to lightest is
squarks $>$ gluinos $>$ $SU(2)$ gauginos $>$ sleptons $>$ bino, then the squark
would usually decay into its corresponding quark and the gluino. This decay can
beat the decay back to jets because of the large QCD coupling and the color
factor associated with the gluino.  A typical whole decay chain yielding final
state leptons would be an extended version of topology B:
\be
\begin{array}{l}
\tilde{b}^c  \to b\, \tilde{g} \\
\phantom{\tilde{b}^c \to b\ }
  \bentarrow 2j\, \chi_2 \\
\phantom{\tilde{b}^c \to b\ \bentarrow 2j\ }
  \bentarrow  \ell\, \tilde{\ell} \\[-2pt] 
\phantom{\tilde{b}^c \to b\ \bentarrow 2j\ \bentarrow \ell\,}
  \bentarrow \ell\, \chi_1 \\ 
\phantom{\tilde{b}^c \to b\ \bentarrow 2j\ \bentarrow \ell\, \bentarrow \ell\ }
   \bentarrow 3j \, ,
\end{array}
\ee
where the LSP decay into 3 jets in the last step proceeds via an off-shell
squark using the $R$-parity violating vertex.  This decay chain of course
assumes that the gluino itself does not decay to 3 jets via the $R$-parity
violating operator.

Even with a SUSY cascade decay, it is not guaranteed to get visible leptons in
the final state.   For example, a superpartner spectrum of the form gluinos $>$
squarks $>$ $SU(2)$  gauginos $>$ sleptons $>$ bino, would typically produce a
much shorter SUSY decay chain, ending with a four jet final state and no leptons. 
This is because the produced squark is an $SU(2)$ singlet which decays directly
into a quark and the bino (the LSP), with the bino eventually decaying to three
quarks.  One can get copious lepton production with a slepton LSP, and the
spectrum  gluinos $>$ squarks $>$ $SU(2)$ gauginos $>$ bino $>$ sleptons would
give
\be
\begin{array}{l}
\tilde{b}^c \to b\, \chi_1 \\
\phantom{\tilde{b}^c \to b\ } \bentarrow \ell\, \tilde{\ell} \\[-2pt]
\phantom{\tilde{b}^c \to b\ \bentarrow \ell\,} \bentarrow \ell\, 3j \, .
\end{array}
\ee
Finally, a spectrum where left- and right-handed sleptons alternate with
neutralinos can even give rise to four leptons in addition to several jets.

\subsection{Resurrecting pair production}
\label{subsec:Resurect_Pair_Prod}

In Sec.~\ref{subsec:nopair}, we argued that QCD pair production of new
colored resonances was not a supermodel.  However, topology C allows for pair
production of new resonances via a supermodel resonance.  

For example, using either a $q\bar{q}$ or a $qq$ resonance, one can produce
vector-like up-type quarks $U$ and $U^c$ with quantum numbers:
\be
\label{eq:stablequarkqnumbers}
\begin{tabular}{c|c|c|c|c}
Field & Spin & $SU(3)_C$ & $SU(2)_L$ & $U(1)_Y$ \\ 
\hline
$U$ & $1/2$ & $\mathbf{3}$ & $-$ & $\phantom{-}2/3$ \\
$U^c$ & $1/2$ & $\mathbf{\overline{3}}$ & $-$ & $-2/3$ \\
\end{tabular}
\ee
They can be produced via the $Z'$ through
\be
g_U\, Z'_\mu \left(\ov{U} \bar\sigma^\mu U 
  - \ov{U^c}\, \bar\sigma^\mu U^c \right) ,
\ee
or via the diquark through
\be
\frac{\tilde{\kappa}_D}{2}\, D\, U^c\, U^c .
\ee

If these new colored particles were exactly stable, they would form
$R$-hadron-like bound states as mentioned above, leading to topology A.   However, the heavy
$U/U^c$ quarks may decay via small CKM-like mixings with the standard model
quarks, leading to
\be
U \to Z + u/c/t \,, \qquad U \to W + d/s/b \,.
\ee
Such decays are not ideal for making a supermodel, since the $W$ ($Z$) boson
only has 22\% (7\%) branching fraction to electrons and muons.   However, if
the $U/U^c$ quarks only couple to other standard model fermions through
higher-dimension operators like 
\be
\label{eq:higherdimopforudecay}
\frac{1}{\Lambda^2}\, (\overline{u^c}\, \sigma^\mu U^c)\,
  (\overline{e^c}\, \sigma_\mu e^c)\,, \qquad 
\frac{1}{\Lambda^3}\, (\overline{d^c}\, \sigma^\mu U^c)\,
  (\overline{\ell}\, h^\dagger \sigma_\mu e^c)\,, 
\ee
then each $U/U^c$ decay can lead to leptons via
\be
U \to e^+ e^- + u/c/t\,, \qquad U \to e^+ \nu + d/s/b\,.
\ee
Operators like Eq.~\Eq{eq:higherdimopforudecay} can always be generated through
sufficiently creative model building, and the scale $\Lambda$ can be made
sufficiently large to evade flavor constraints while still being small enough to
give prompt decay.

Another option to force leptons to appear in the final state is to have a resonance decay to pairs of
colored particles that also carry lepton number, such as leptoquarks or even
the leptodiquarks of Eq.~\Eq{eq:leptodiqnumbers2}.\footnote{To produce such states
from the diquark resonance, one needs two different fields of opposite
lepton number but same baryon number, i.e.\ a leptoquark and an antilepto-quark.}
Alternatively, through a  $q\bar{q}$ resonance, one can pair produce color singlet objects with lepton number,
such as ``sleptons'' \cite{Baumgart:2006pa}, as long as the $q\bar{q}$ resonance
does not couple to standard model leptons directly.

Finally, a neutral $q\bar{q}$ resonance can dominantly decay to two secondary
resonances that carry no standard model charges. These secondary resonances have
a near infinity of possible decay modes, opening up a huge range of final state
possibilities.  Such scenarios will be supermodels as long as the secondary
resonances have an $\mathcal{O}(1)$ branching fraction to highly visible final
states.

\section{Conclusions}
\label{sec:conclude}

In this paper, we investigated some new physics scenarios which could be probed
by  a low energy and low luminosity initial LHC data set, and which will not have
been ruled out by the Tevatron and other measurements.  We call such scenarios
{\em supermodels\/}; they are not necessarily motivated by usual model building
goals such as solving the hierarchy problem, but are constructed to demonstrate
that high production cross sections and clean experimental signatures are
possible for early LHC.

Assuming perturbative couplings, we found that $s$-channel production of $qq$ or
$q\bar{q}$ resonances are the most promising supermodel scenarios. To supersede
the Tevatron sensitivities in searches for pair-produced particles or single
resonances produced from an initial state involving gluon(s), the LHC typically
needs higher integrated luminosity than considered in this paper.  Not
unexpectedly, given a $pp$ collider, we found that resonances that couple to
$qq$ initial states have by far the largest LHC cross sections, and in this
channel there is a large space of supermodels (see
Fig.~\ref{fig:twowedgeplots}).

We explored various decay topologies of the produced resonances that lead to
easily identifiable final states containing a pair of charged leptons or other
(quasi-)stable charged particles.  While the possibilities for such decays are only
limited by theorists imagination and model building ingenuity, we presented some
simple examples focusing on decay topologies that also arise in more standard
extensions of the standard model. While the supermodels exhibited here might
not be as attractive as the name suggests, many of the same final state signatures
are useful search channels for ``well motivated'' models as well. 
Therefore, searching for supermodels with the early LHC data may benefit finding
prettier models when larger data samples become available. 

Finally, the space of interesting early LHC searches would be extended if (i)
nonperturbative couplings are considered; (ii) pair production is enhanced by
high particle multiplicities; (iii) one compares to the currently published
Tevatron bounds (some of which utilize less than 100 pb$^{-1}$ of data) instead
of the 2010 Tevatron sensitivity with 10~fb$^{-1}$ of data; (iv) the early LHC
data used for analysis approaches or goes beyond 100~pb$^{-1}$.

\begin{acknowledgments}

We thank Mina Arvanitaki, Beate Heinemann, and Matthew Schwartz for helpful conversations.
This work was supported in part by the Director, Office of Science, Office of
High Energy Physics of the U.S.\ Department of Energy under contract
DE-AC02-05CH11231.  M.S. is supported by DE-FG02-01ER-40676.
J.T.\ acknowledges support from the Miller Institute for Basic Research in Science.
D.W.\ was supported by a University of California Presidential Fellowship.  M.S. and J.T.  thank
the Aspen Center for Physics for their hospitality while this work was in preparation. 

\end{acknowledgments}

\end{document}